\documentclass{nature}
\usepackage{}
\usepackage{amssymb}
\usepackage{amsmath}
\usepackage{amsfonts}
\usepackage{graphicx}
\usepackage{caption}
\usepackage{subfig}
\usepackage{txfonts}
\usepackage{epsfig}
\begin{document}

\title{Layer Anti-Ferromagnetism on Bilayer Honeycomb Lattice}
\author{Hong-Shuai Tao, Yao-Hua Chen, Heng-Fu Lin, Hai-Di Liu and Wu-Ming Liu$^\star$}

\maketitle

\begin{affiliations}
\item
Beijing National Laboratory for Condensed Matter Physics,
Institute of Physics, Chinese Academy of Sciences,
Beijing 100190, China

$^\star$e-mail: wliu@iphy.ac.cn

\end{affiliations}

\begin{abstract}
Bilayer honeycomb lattice, with inter-layer tunneling energy, has a parabolic dispersion relation, which causes the charge imbalance between two sublattices. Here, we investigate the metal-insulator and magnetic phase transitions on the strongly correlated bilayer honeycomb lattice by cellular dynamical mean-field theory combined with continuous time quantum Monte Carlo method. We find that different kinds of magnetic spontaneous symmetry breaking on dimer and non-dimer sites, cause a novel phase transition between normal anti-ferromagnet and layer anti-ferromagnet. We sketch the phase diagrams as the function of temperature, interaction and inter-layer hopping. Finally, we set up an experimental protocol for cold atoms in optical lattice to observe these phenomena in future experiments.
\end{abstract}

Bilayer honeycomb lattice (BHL) has attracted enormous interest in both experimental and theoretical research. Lots of novel phenomena have been found in BHL, for instance, the quantum Hall effect, quantum spin Hall effect, and chiral superconductivity \cite{Kharitonov, Novoselov, Freitag, Feldman, Zhao, Maher, Weitz, Kane, Hosseini}. However, the charge and magnetic order induced by Coulomb interaction are still challenge in the strongly correlated BHL \cite{Zhang, Mezzacapo, Anderson, Wang, Wu}. In BHL, a quadratic dispersion relation signed by two touching bands in the corners of Brillouin zone, which is driven by the inter layer hopping. In addition, the spontaneous symmetry can be broken by the dimers when the inter-layer hopping changes. Some amazing phases emerge, such as layer anti-ferromagnetic phase and paramagnetic insulator phase. Previous work mainly focus on the electronic properties of BHL \cite{Vafek, McCann, Nilsson, McCann2, Nilsson2, Lopes, Abergel}. However, the interesting magnetic phases induced by the inter-layer hopping and Coulomb interaction are absent. Be different to mono-layer honeycomb lattice, different asymmetry in dimer and non-dimer sites promise a more exciting phase diagram. The progress of optical lattice provides us a useful tool to set a controllable and clearness experimental platform to simulate the strongly correlated BHL, in which the interaction between trapped fermionic cold atoms can be tuned by the Feshbach resonance \cite{Jaksch, Hofstetter, Greiner, Duan, Soltan, Gemelke, Tao}. Up to now, mono-layer honeycomb lattice and bilayer graphene have been intensively studied either experimentally or numerically, however in bilayer honeycomb system with strongly correlated interaction no comprehensive conclusion has been achieved.

To deal with strongly correlated systems, dynamical mean-field theory (DMFT) has been proved to be a very useful and effective tool \cite{Metzner, Georges1, Bulla, Georges2}, which has made significantly progress in the field of metal-insulator transition. In the infinite-dimensional limit, it is exact that the self-energy independent of momentum. However, in low-dimensional systems, the quantum fluctuation and short range correlations play an important role, which are ignored in DMFT. The cellular DMFT (CDMFT), as a cluster extension of DMFT, effectually incorporates the spatial correlations by mapping the many-body problem into local degrees of freedom treated exactly within a finite cluster that is embedded in a self-consistent bath \cite{Kotliar, Maier, Tong}. In two-dimensional systems, quantum fluctuations are strong, CDMFT will be more precise than DMFT, which is more effective to investigate the phase transition in low and multi-component systems \cite{Bolech, congjun, Tao}.

In this report, we investigate the finite temperature metal-insulator and magnetic phase transition in strongly correlated bilayer honeycomb lattice (BHL). We improve cellular dynamical mean-field theory (CDMFT) combined with continue-time quantum Monte Carlo (CTQMC) method \cite{Rubtsov}, which is used as impurity solver. By investigating the density of states (DOS) and magnetization, we find a phase transition from paramagnetic phase to anti-ferromagnetic phase. In a proper value of inter-layer hopping, a novel layer anti-ferromagnetic phase will emerge. As we keep on increasing the inter-layer hopping, the layer anti-ferromagnetic phase will transform to a paramagnetic phase. In this system, the nonlocal inter-layer hopping plays an important role on localizing the free election and modifying the spatial distribution of the electron in lattice sites, especially in dimer sites. We have presented the DOS, double occupancy, and fermi surface below, which can be directly detected in future experiments.

\section*{Results}
\subsection{The strongly correlated bilayer honeycomb lattice.}
As shown in Fig. 1 (a1), the sublattice on top-layer is signed by $A_1 (B_1)$, and $A_2 (B_2)$ denotes the sublattice on bottom. $A_1$ and $A_2$ are connected by the inter-layer bands. The density of states (DOS) for different sublattices at $U=0$ and $t_1=t$ is shown in Fig. 1 (c), which is much different from the mono-layer honeycomb lattice \cite{Zhu}. There is no band gap between the conduction and valence bands in BHL. The low-energy dispersion is quadratic, which is linear in mono-layer case. For this lattice structure, we consider the standard Hubbard model:\begin{eqnarray}
&&H=-t\sum_{\langle ij\rangle \sigma\alpha}(c^{\dag}_{i\sigma\alpha}c_{j\sigma\alpha}+h.c.)+U\sum_{i\alpha}n_{i\downarrow\alpha}n_{i\uparrow\alpha}-t_1\sum_{i\sigma\alpha}c^{\dag}_{i\sigma\alpha}c_{i\sigma(1-\alpha)},
\end{eqnarray}where $c^\dag_{i\sigma\alpha}$ ($c_{i\sigma\alpha}$) denotes the creation (annihilation) operator of fermionic atoms on site $i$ with spin $\sigma$, and layer parameter $\alpha$ ($\alpha=0$ in the top-layer and $\alpha=1$ in the bottom-layer), $n_{i\sigma}=c^\dag_{i\sigma\alpha}c_{i\sigma\alpha}$ is the density operator. $U$ is the on-site Coulomb repulsion. The intra-layer nearest neighbor hopping is $t$, and $t_1$ is the inter-layer hopping. In this report, we set $t$ as energy unit ($t=1$).

\subsection{The metal-insulator phase transition. }
The phase diagram obtained at $T/t=0.1$ shows that the crucial line of Single particle excitation gap and magnetization, at half filled, divide the phase into four regions. As shown in Fig. 2, when $2<t_1/t<4.8$, the system stays anti-ferromagnet. As we increasing interaction $U$, the system undergoes from anti-ferromagnetic metal to anti-ferromagnetic insulator, and the transition line is denoted as red solid line. When $t_1/t<2$, magnetic phase transition will occur, as we increase $U$, with the indication of blue solid line. In the value of $0<t_1/t<0.2$, when $U>U_c$ ($U_c/t=4.7$, at $t_1/t=0$), we will get a small region, which is named as paramagnetic insulator (PI). It is the candidate of quantum spin liquid for its non-magnetic insulating property and being similar with resonating valence bonds state\cite{Meng, Hohenadler, Rachel}.

Double occupancy always be used to measure the localization of the electrons directly and indicates the transition order, which is an important parameter to the crucial point \cite{Kancharla}. In our paper, we investigate the double occupancy $D_{occ}=\partial{F}/\partial{U}=\frac{1}{4}\sum_i<n_{i\uparrow}n_{i\downarrow}>$ as the function of inter-layer hopping $t_1$ for various interaction $U$ [see in Fig. 3], in which $F$ is free energy. The $D_{occ}$ decreases continuously when interaction $U$ increase for $t_1/t=0$. This suggests the interaction enhances the localization of electrons and induces an insulating state in the system. We use $D_{AA}$ to describe the $D_{occ}$ for $A_1$ and $A_2$ sites (hollow circles in Fig. 3), and the $D_{occ}$ for $B_1$ and $B_2$ sites are signed as $D_{BB}$ (solid circles in Fig. 3). $D_{AA}$ is separated from $D_{BB}$, when $t_1\neq t$. The $D_{AA}$ increases while the $D_{BB}$ decreases as $t_1$ increases. This suggests the itinerancy of electrons in dimer sites is enhanced due to the increasing inter-layer hopping. Be different with $D_{AA}$, $D_{BB}$ decreases while $t_1$ increases. This result suggests the intra-hopping between dimer and non-dimer sites is weaken due to forming of spin-polarized electrons in dimer sites. A second-order phase transition can be confirmed by the linear and smooth developing $D_{AA}$ and $D_{BB}$ even in critical points.

In order to find the metal-insulator phase transition as the evolution of single particle spectral \cite{Hu}, we define DOS, which has the form like \begin{eqnarray}
D(\omega)=-\frac{1}{\pi}\sum^{12}_{i=1}(ImG_{ii}(\omega-i\delta)),
\end{eqnarray}
where $i$ denotes the sublattice index. The DOS can be derived from the imaginary time Green's function $G(\tau)$, which is obtained by maximum entropy method \cite{Jarrell}. Fig. 4 (a) shows the DOS for different inter-layer hopping when $U/t=2.5$, $T/t=0.1$. It is found that, in weak interaction, the inter-layer hopping $t_1$ does not affect the metallic properties of BHL. In Fig. 4 (b), we can find that the system keeps at a metallic state when $t_1/t=1.8$, and an obvious pseudo-gap is formed when $t_1/t=2.2$. A metal-insulator transition happens when $t_1/t=3.2$ for $U/t=3.5$ and $T/t=0.1$. At large interaction, such as $U/t=6.0$ [see Fig. 4 (c)], the system stays insulating phase, which is insensitive of $t_1$. As shown in Fig. 4 (d), we fix the inter-layer hopping $t_1/t=1.0$ and temperature $T/t=0.1$, and find that the metal-insulator phase transition occurs as we increase the interaction $U$. The procedure will be the same as Fig. 4 (c), which remind us that inter-layer hopping and Coulomb interaction play the same role for the metal-insulator transition, at intermediate value of $U$ and low temperature $T/t\approx0.1$.

\subsection{The magnetic phase transition and a novel layer anti-ferromagnetic phase.}In strongly correlated BHL, charge imbalance between the two sublattices sites causes different kinds magnetic spontaneous symmetry breaking, which divides the sites into dimer sites and non-dimer sites. The parameter $<n_{i\sigma}>$ indicates the electron density in lattice site $i$ with spin index $\sigma$. In order to find the magnetic order formed in BHL, we use a magnetic order parameter defined as $m=\frac{1}{N_c}\sum_i(<n_{i\uparrow}>- <n_{i\downarrow}>)$, where $i$ denotes the lattice index belongs to different sublattices such as $A_1, A_2, B_1$ and $B_2$ [see in Fig. 1 (a1)]. We set magnetization of $A_1$ as positive sign. Fig. 6 shows the evolution of $m$ as a function of $t_1$ for different $U$. We can find that when $U/t=4.0$, the system keeps at a paramagnetic state in weak $t_1$. A magnetic state with anti-ferromagnetic order is found when $t_1/t=1.0$. $m$ decrease continuously when $t_1/t>1.0$, and the magnetic of $A_1/A_2$ decreases to zero while $B_1/B_2$ keeps nonzero. This novel phase is called layer anti-ferromagnetic insulator, in which A sites exhibit a paramagnetic property and B sites are anti-ferromagnetic ordered. The layer anti-ferromagnetic insulator transform to a paramagnetic insulator when $t_1/t>4.6$. Sketches of the possible magnetic order existing in the BHL is show in Fig. 7 (a).

Finally, phase diagram of magnetization about $t_1$ and $U$ at $T/t=0.1$ is shown in Fig. 7. In weak $U$ case ($U/t<4.7$), system transforms from paramagnet to anti-ferromagnet. A layer anti-ferromagnetic phase is found at large $t_1$, in which the magnetic of dimer sites keeps zero while non-dimer sites is nonzero. As we keep increasing the inter-layer hopping $t_1$, the system can transform from the layer anti-ferromagnetic state to paramagnetic state.

\subsection{Experimental protocol.}We propose an experiment setup to investigate the phase transition in strongly correlated bilayer honeycomb lattice (BHL). The ${}^{40}$K atoms can be produced as a pure fermion condensate by evaporative cooling \cite{O'Hara} , which provides two hyperfine states $|F, m_F\rangle=|9/2, -9/2\rangle\equiv|\uparrow\rangle$ and $|F, m_F\rangle=|9/2, -7/2\rangle\equiv|\uparrow\rangle$ \cite{Lucia}. Three standing-wave laser beams are used to form the honeycomb lattice, and two extra laser beams along the $z$ direction suppress the tunneling between layers \cite{Duan}. The potential of optical lattice is given by $V_{h}(x,y)=V_0\sum_{j=1,2,3}sin^2[k(x\cos\theta_j+y\sin\theta_j)+\pi/2]$, where $\theta_1=\pi/3, \theta_2=2\pi/3, \theta_3=0$. Then, we use another three standing-wave laser beams with a $2\pi/3$ angle between each other to form triangular lattice. The potential is given by $V_t(x, y)=V_0[3+4\cos(k_xx/2)\cos(\sqrt{3}k_yy/2)+2\cos(\sqrt{3}k_yy)]$. $k_x$ and $k_y$ are the two components of the wave vector $k=2\pi/\lambda$ in these two types of lattices, where $\lambda=738nm$ is wavelength of the laser, and $V_0$ is given in recoil energy $E_r=\hbar^2k^2/2m$. Inserting the triangular lattice between two layers of honeycomb lattice, the Bernal stacking BHL with trapped ${}^{40}$K atoms will be formed \cite{Tung,Hou}. In BHL the intra-layer hopping $t=(4/\sqrt{\pi})E_r^{1/4}V_0^{3/4}exp[-2(V_0/E_r)^{1/2}]$ is adjusted by the periodic potential of laser beam and $t_1$ can be tuned by changing the wavelength of laser beam in $z$ direction. The on-site interaction $U=\sqrt{8/\pi}ka_sE_r(V_0/E_r)^{3/4}$ determined by the s-wave scattering length $a_s$, which can be tuned by Feshbach resonance, and the temperature can be extracted from the time-of-flight images \cite{Schneider}.

It should be mentioned that, in cold atom optical lattice, to observed the double occupied sites, firstly we have to increase the depth of the optical lattice to prevent further tunneling of atoms. Next, we shift the energy of the atoms on doubly occupied sites by approaching a Feshbach resonance. Then the one spin component of atoms on double occupied sites will transfer to a new magnetic sublevel by radio-frequency pulse method. Finally, we will deduce the double occupancy by the absorption images \cite{Jordans,stoferle}.

To get Fermi surface in experimental, we ramp down the optical lattice slowly enough, and the atoms will stay adiabatically in the lowest band while quasi-momentum is approximately conserved. We lower the lattice potential to zero rapidly, after that we switch off the confining potential and ballistic expand for several milliseconds. Then we take an absorption images, which is the Fermi surface \cite{Kohl,Chin}.

\section*{Discussion}
In this work, we have investigated the metal-insulator transition and magnetic phase transition in strongly correlated bilayer honeycomb lattice using cellular dynamical mean-field theory (CDMFT) combining with continue-time quantum Monte Carlo (CTQMC) method. In low-energy cases, we map the phase diagram as a function of interaction $U$, inter-layer hopping $t_1$ and magnetization $m$. It shows that the inter-layer hopping affects the electrons to form spin-polarized electrons, and an insulating state is induced. A layer anti-ferromagnetic phase is found at large $t_1$, in which the magnetization of dimer sites is zero while non-dimer keeps finite. Therefore, the inter-layer hopping $t_1$ plays an important role to form a singular magnetic spontaneous symmetry breaking phase. Our study may provide a helpful step for understanding the interaction and inter-layer hopping driven metal-insulator transition, the exotic magnetic order with asymmetry and nature of CDMFT method used in bilayer honeycomb lattice.

\section*{Methods}
\subsection{The cellular dynamical mean-field theory.}
We combine the cellular dynamical mean-field theory (CDMFT) with continuous time quantum Monte Carlo (CTQMC) method to determine the metal-insulator transition and magnetic phase transition in the strongly correlated bilayer honeycomb lattice. In low-dimensional systems, quantum fluctuations are much stronger than the higher dimensions. The nonlocal effect will be much important in this case. Dynamical mean-field theory ignoring the nonlocal correlations will lead lots of errors in calculation. Therefore, we use CDMFT, as the advanced method in our work. We map the original lattice onto a 12-site effective cluster embedded in a self-consistent bath field [see Fig. 1 (a2)]. Starting with a guessing self-energy $\Sigma(i\omega)$ (which is independent of momentum \cite{Hartmann}), we can get the Weiss field $G_0(i\omega)$ obtained by the coarse-grained Dyson equation:
\begin{eqnarray}
G^{-1}_0(i\omega)=(\sum_{\textbf{K}}\frac{1}{i\omega-t(\textbf{K})-\Sigma(i\omega)})^{-1}+\Sigma(i\omega),
\end{eqnarray}
where $\omega$ is Matsubara frequency, $\mu$ is the chemical potential, $\textbf{K}$ is in the reduced Brillouin zone of the super-lattice, and $t(\textbf{K})$ is hopping matrix for the super-lattice. The form of $t(\textbf{K})$ is:
\begin{eqnarray}
t(\textbf{K})=\left(
\begin{array}{cccccccccccc}
    0 & t & 0 & t\delta_1 & 0 & t & 0 & 0 & 0 & t_1\delta_1 & 0 & 0\\
    t & 0 & t & 0 & t\delta_2 & 0 & 0 & 0 & 0 & 0 & 0 & 0\\
    0 & t & 0 & t & 0 & t\delta_3 & 0 & t_1 & 0 & 0 & 0 & 0\\
    t\delta_1^* & 0 & t & 0 & t & 0 & 0 & 0 & 0 & 0 & 0 & 0\\
    0 & t\delta_2^* & 0 & t & 0 & t & 0 & 0 & 0 & 0 & 0 & t_1\\
    t & 0 & t\delta_3^* & 0 & t & 0 & 0 & 0 & 0 & 0 & 0 & 0\\
    0 & 0 & 0 & 0 & 0 & 0 & 0 & t & 0 & t\delta_1 & 0 & t\\
    0 & 0 & t_1 & 0 & 0 & 0 & t & 0 & t & 0 & t\delta_2 & 0\\
    0 & 0 & 0 & 0 & 0 & 0 & 0 & t & 0 & t & 0 & t\delta_3\\
    t_1\delta_1^* & 0 & 0 & 0 & 0 & 0 & t\delta_1^* & 0 & t & 0 & t & 0\\
    0 & 0 & 0 & 0 & 0 & 0 & 0 & t\delta_2^* & 0 & t & 0 & t\\
    0 & 0 & 0 & 0 & t_1 & 0 & t & 0 & t\delta_3^* & 0 & t & 0\\
\end{array}
\right),
\end{eqnarray}
where $\delta_1=e^{i\textbf{K}\cdot\textbf{a}_1}$, $\delta_2=e^{i\textbf{K}\cdot(\textbf{a}_1-\textbf{a}_2)}$, $\delta_3=e^{-i\textbf{K}\cdot\textbf{a}_2}$ and $\textbf{a}_1$, $\textbf{a}_2$ are real lattice vectors as shown in Fig. 1 (a1).
The cluster Green's function $G(i\omega)$ can be gotten by the impurity solver. In our work, we use the numerically exact CTQMC simulation as impurity solver and take $5\times10^6$ QMC sweeps for each CDMFT loop \cite{Rubtsov}. The new self-energy $\Sigma(i\omega)$ is recalculated by the Dyson equation:
\begin{eqnarray}
\Sigma(i\omega)=G^{-1}_0(i\omega)-G^{-1}(i\omega).
\end{eqnarray}
This iterative loop repeated until self-energy is converged.

The CTQMC method as impurity solver can be taken as follows. We start the procedure at partition function, which can be written as:
\begin{eqnarray}
Z=T_re^{-\beta H}=Z_0T_{\tau}[\sum_k \frac{1}{k!}(-\int_0^{\beta}H_1(\tau)d\tau)^k],
\end{eqnarray}
where $T_{\tau}$ is time-ordering operator, $H_1(\tau)=e^{\tau H_0}H_1e^{-\tau H_0}$ is $H_1$ in the interaction picture, and $Z_0=T_re^{-\beta H_0}$ is a partition function for the unperturbed term. Putting $H_1=U\sum_in_{i\downarrow}n_{i\uparrow}$ in Eq. 6, the partition function will be
\begin{eqnarray}
Z=Z_0\sum_k\frac{(-U)^k}{k!}\int\cdot\cdot\cdot\int dr_1\cdot\cdot\cdot dr_k\langle T_{\tau}n_{\uparrow}(r_1)\cdot\cdot\cdot n_{\uparrow}(r_k)\rangle_0\langle T_{\tau}n_{\downarrow}(r_1)\cdot\cdot\cdot n_{\downarrow}(r_k)\rangle_0.
\end{eqnarray}
Here $\langle\rangle_0$ indicates a theromdynamic average with respect to $e^{-\beta H_0}$. Using Wick's theorem, for each order in $k$, $\langle T_{\tau}n_{\sigma}(r_1)\cdot\cdot\cdot n_{\sigma}(r_k)\rangle_0$ ($\sigma=\uparrow, \downarrow$) can be written as determinant $detD(k)$:
\begin{eqnarray}
D(k)=\left(
\begin{array}{cccc}
G^0(r_1, r_1) & G^0(r_1, r_2) & \cdots & G^0(r_1, r_k)\\
G^0(r_2, r_1) & G^0(r_2, r_2) & \cdots & G^0(r_2, r_k)\\
\cdot & \cdot & \cdot & \cdot\\
\cdot & \cdot & \cdot & \cdot\\
G^0(r_k, r_1) & G^0(r_k, r_2) & \cdots & G^0(r_k, r_k)\\
\end{array}
\right),
\end{eqnarray}
where $G^0$ is non-interacting Green's function. There is no spin index in $D(k)$ for the determinants of spin-un and -down being equivalent. Like classical Monte Carlo, by integrand of Eq. 7, we can get the weight of order $k$
\begin{eqnarray}
W_k=(-\delta\tau U)^kdetD_{\uparrow}(k)detD_{\downarrow}(k),
\end{eqnarray}
where $\delta\tau=\beta/L$ is slice of imaginary time. We can get the standard Metropolis acceptance ratio $R$ of adding vertex by the detailed balance condition:
\begin{eqnarray}
\frac{1}{L\cdot N}W_kP_{k\rightarrow k+1}=\frac{1}{k+1}W_{k+1}P_{k+1\rightarrow k},
\end{eqnarray}
\begin{eqnarray}
R=\frac{P_{k\rightarrow k+1}}{P_{k+1\rightarrow k}}=-\frac{U\beta N}{k+1}\left(\frac{detD_{\uparrow}(k+1)detD_{\downarrow}(k+1)}{detD_{\uparrow}(k)detD_{\downarrow}(k)}\right).
\end{eqnarray}
Here $P_{k\rightarrow k+1}$ is the probability to increase the order from $k$ to $k+1$ ($P_{k+1\rightarrow k}$ the probability to decrease the order from $k+1$ to $k$), $\frac{1}{L\cdot N}$ is probability to choose a position in time and space for vertex you intend to add while $\frac{1}{k+1}$ is the probability to choose one vertex you intend to remove of from the existing $k+1$ noes. To calculate the ratio $R$, we have to deal with the function $detD(k+1)/detD(k)$.
\begin{eqnarray}
detD(k+1)/detD(k)=det(I+(D(k+1)-D(k))M(k))=\lambda,
\end{eqnarray}
$M_{\sigma}(k)=D_{\sigma}^{-1}(k)$, we can easily get the value of $\lambda$ in matrix form:
\begin{eqnarray}
&&det\left(
\begin{array}{ccccc}
1 & 0 & \cdots & 0 & G^0(r_1, r_{k+1})\\
0 & 1 & \cdots & 0 & G^0(r_2, r_{k+1})\\
\cdot & \cdot & \cdots & \cdot & \cdot\\
\cdot & \cdot & \cdots & \cdot & \cdot\\
0 & 0 & \cdots & 1 & G^0(r_k, r_{k+1})\\
G^0(r_{k+1}, r_i)M(k)_{i, 1} & G^0(r_{k+1}, r_i)M(k)_{i, 2} & \cdots & G^0(r_{k+1}, r_i)M(k)_{i, k} & G^0(r_{k+1}, r_{k+1})\\
\end{array}
\right)\nonumber\\
&&=G^0(r_{k+1}, r_{k+1})-G^0(r_{k+1}, r_i)M(k)_{i, j}G^0(r_j, r_{k+1})=\lambda.
\end{eqnarray}
Then it is easy to obtain the update $M$ for the order $k+1$ by numerical method:
\begin{eqnarray}
M(k+1)=\left(
\begin{array}{cccc}
\cdot & \cdot & \cdot & -L_{1, k+1}\lambda^{-1}\\
\cdot & M'_{i, j} & \cdot & -L_{2, k+1}\lambda^{-1}\\
\cdot & \cdot & \cdot & -L_{k, k+1}\lambda^{-1}\\
\cdot & \cdot & \cdot & \cdot\\
-\lambda^{-1}R_{k+1, 1} & -\lambda^{-1}R_{k+1, 2} & \cdots & -\lambda^{-1}\\
\end{array}
\right),
\end{eqnarray}
where the factor of the matrix is $M'_{i, j}=M(k)_{i, j}+L_{i, k+1}\lambda^{-1}R_{k+1, j}$, $R_{i, j}=G^0(i,l)M(k)_{l, j}$ and $L_{i, j}=M(k)_{i, l}G^0(l, j)$. For the step $k-1$, we can also get the radio $R$ and update formulas of $M(k-1)$:
\begin{eqnarray}
&&R=-\frac{k}{U\beta N}\left(\frac{detD_{\uparrow}(k-1)detD_{\downarrow}(k-1)}{detD_{\uparrow}(k)detD_{\downarrow}(k)}\right),\\
&&M_{i, j}(k-1)=M_{i, j}(k)-M_{i, l}(k)M_{l, j}(k)/M_{l, l}(k).
\end{eqnarray}
Using the update formula for $M$, the Green's function can be obtained both in imaginary time and at Matsubara frequencies:
\begin{eqnarray}
G(\tau-\tau')=G^0(\tau-\tau')-G^0(\tau-\tau_i)M_{i, j}G^0(\tau_j-\tau'),\nonumber\\
G(i\omega)=G_0(i\omega)-G_0(i\omega)\left[ \frac{1}{\beta}\sum_{i, j}M_{i, j}e^{-\omega(\tau_i-\tau_j)}\right]G_0(i\omega).
\end{eqnarray}
Here $G_0(i\omega)$ is a bare Green's function.


\begin{addendum}

\item [Acknowledgement]
This work was supported by the NKBRSFC under grants Nos. 2011CB921502, 2012CB821305, and NSFC under grants Nos. 61227902, 61378017.

\item [Author Contributions]
H. S. T. performed calculations.
H. S. T., Y. H. C., H. D. L, H. F. L., W. M. L. analyzed numerical results.
H. S. T., Y. H. C., W. M. L. contributed in completing the paper.

\item [Competing Interests]
The authors declare that they have no competing financial interests.

\item [Correspondence]
Correspondence and requests for materials should be addressed to Hong-Shuai Tao and Wu-Ming Liu.

\end{addendum}

\clearpage

\newpage
\bigskip

\textbf{Figure 1 The structure of bilayer honeycomb lattice and its qualities in the non-interacting limit.}
(a1): Bernal stacking of the bilayer honeycomb in real-space with intra- and inter-layer hopping $t$ and $t_1$ between the sublattice $A_1$, $B_1$ on top-layer and $A_2$, $B_2$ on bottom-layer. The black arrows $a_1$ and $a_2$ are the lattice vectors. (a2): The cellular of our bilayer system in cellular dynamical mean-field theory (CDMFT). The red solid line containing sites 0, 1, 2, 3, 4, 5 belongs to top-layer and blue dotted line with sites 6, 7, 8, 9, 10, 11 belongs to bottom-layer. (b): Reciprocal lattice of bilayer honeycomb lattice with $b_1$ and $b_2$ being reciprocal lattice vectors. The thick red line shows the first Brillouin zone. The $\Gamma$, $K, M$ and $K'$ points denote the points with different symmetry in first Brillouin zone. (c): Density of states of our system for $A_1$/$A_2$ and $B_1$/$B_2$ sites where $U=0$ at half filling.

\textbf{Figure 2 Metal-insulator Phase diagram in fixed $T$ or $t_1$.}
Phase diagram as a function of inter-layer hopping $t_1$ and interaction $U$ at $T/t=0.1$. The red solid line shows the metal-insulator phase transition, and blue solid line denotes the staggered magnetization $M=n_{\uparrow}-n_{\downarrow}$, which divides the phase into paramagnetic metal (PM), paramagnetic insulator (PI), anti-ferromagnetic metal (AFM) and anti-ferromagnetic insulator (AFI). Inset: Phase diagram as a function of temperature $T$ and interaction $U$ at fixed $t_1$.

\bigskip

\textbf{Figure 3 The evolution of double occupancy $D_{occ}$.}
The double occupancy as a function of inter-layer hopping $t_1$ for different interaction $U$ at temperature $T/t=0.1$. The dark blue arrow at $t_1/t=1.2$ denotes the phase transition point at $U/t=4.5$. The dimer sites tend to be double occupied however non-dimer sites tend to be single occupied, with increasing $t_1$.

\bigskip

\textbf{Figure 4 The density of states.}
(a), (b), (c): The density of states as a function of frequency $\omega$ for different inter-layer hopping $t_1$ at temperature $T/t=0.1$. (a): The metallic phase at $U/t=2.5$. (b): As we increase $t_1$, system undergoes a phase transition from metal (at $U/t=3.5$ and $t_1/t=1.8$) to insulator (at $t_1/t=3.2$). Single particle excitation gap will open at about $t_1/t=2.4$. (c): The insulating phase at $U/t=6.0$. There is a visible single particle excitation gap around the Fermi energy. (d): The density of states as a function of $\omega$ for different $U$ with fixed $t_1$.

\bigskip

\textbf{Figure 5 The evolution of Fermi surface.}
The energy spectral as a function of momentum $k$ for different inter-layer hopping $t_1$ at $T/t=0.1$ and fixed $U/t=3.5$: (1a) $t_1/t=0.0$, (1b) $t_1/t=1.0$, (1c) $t_1/t=4.0$. The Fermi surface as a function of $k$ for different interaction at $T/t=0.1$ and fixed $t_1/t=1.2$: (2a) $U/t=2.0$, (2b) $U/t=2.8$, (2c) $U/t=5.0$.

\bigskip

\textbf{Figure 6 The evolution of the magnetic order parameter $m$.}
The evolution of magnetic order parameter $m$ at $T/t=0.1$ and $U/t=4.0$. A paramagnetic phase occur in weak $t_1$. For $t_1/t>1.0$, the magnetic parameter $m$ is nonzero and has opposite sign between $A_1/A_2$ sites and $B_1/B_2$ sites. The system goes into anti-ferromagnetic phase. At large $t_1$ the magnetic of $A_1/A_2$ sites are more easily decreasing to zero while $B_1/B_2$ sites are still nonzero. The system will be layer anti-ferromagnetic phase. Single particle excitation gap $\Delta$E denoted by the dark green solid line, divides the phase into paramagnetic metal (PM), paramagnetic insulator (PI) and anti-ferromagnetic insulator (AFI).

\bigskip

\textbf{Figure 7 The phase diagram of magnetic phase transition.}
In weak interaction $U$ and weak inter-layer hopping $t_1$, the system will be paramagnetic phase. When we increase $U$, the system will undergo a magnetic phase transition to anti-ferromagnetic phase. When we increase $t_1$, the magnetization of $A_1/A_2$ sites decrease to zero while $B_1/B_2$ sites stay nonzero. The system goes to layer anti-ferromagnetic phase. In the region where $t_1/t>5.0$, the system returns to paramagnetic phase.

\newpage
\begin{figure}
\begin{center}
\epsfig{file=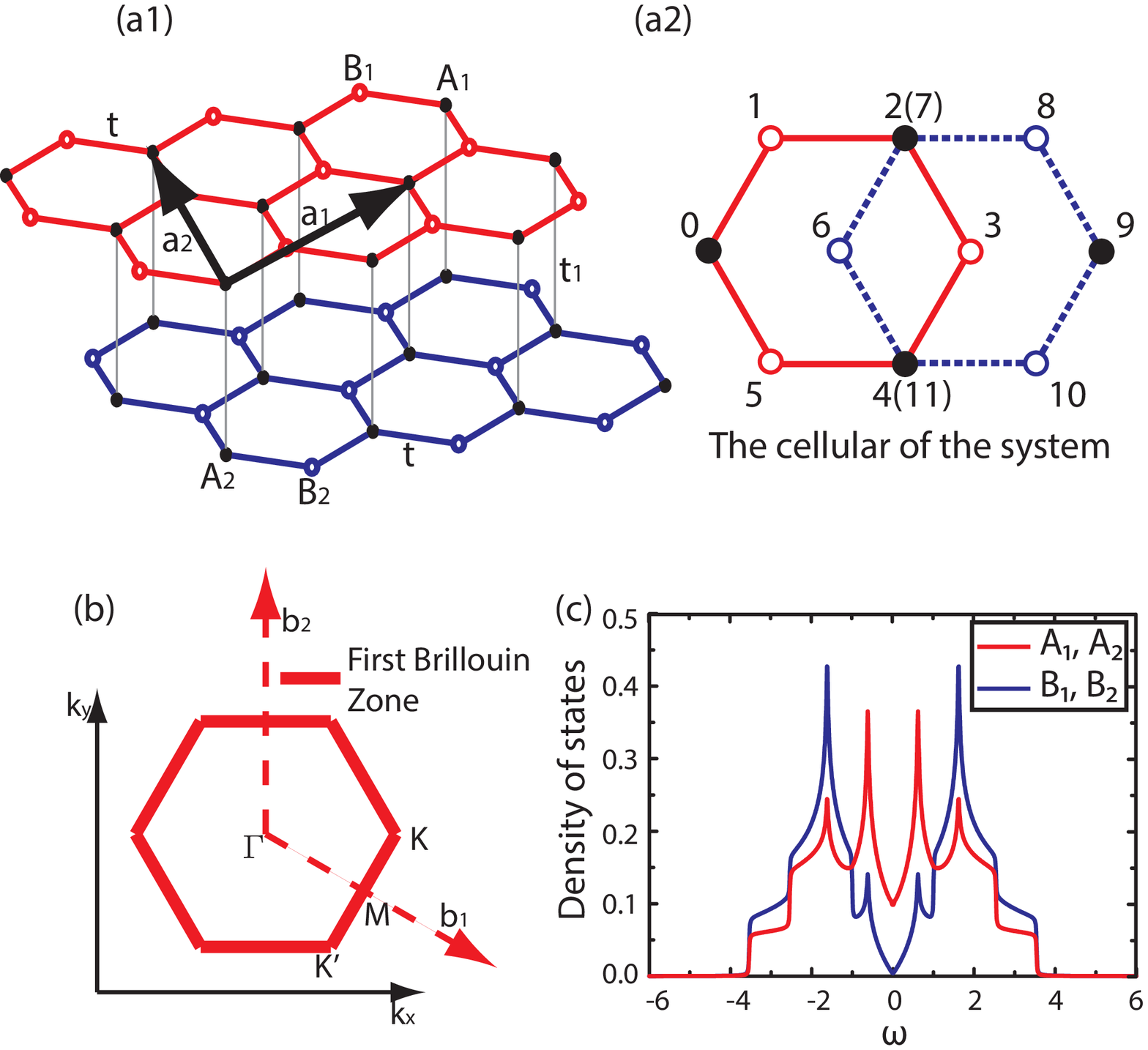,width=15cm}
\end{center}
\label{fig:Lattice}
\end{figure}

\begin{figure}
\begin{center}
\epsfig{file=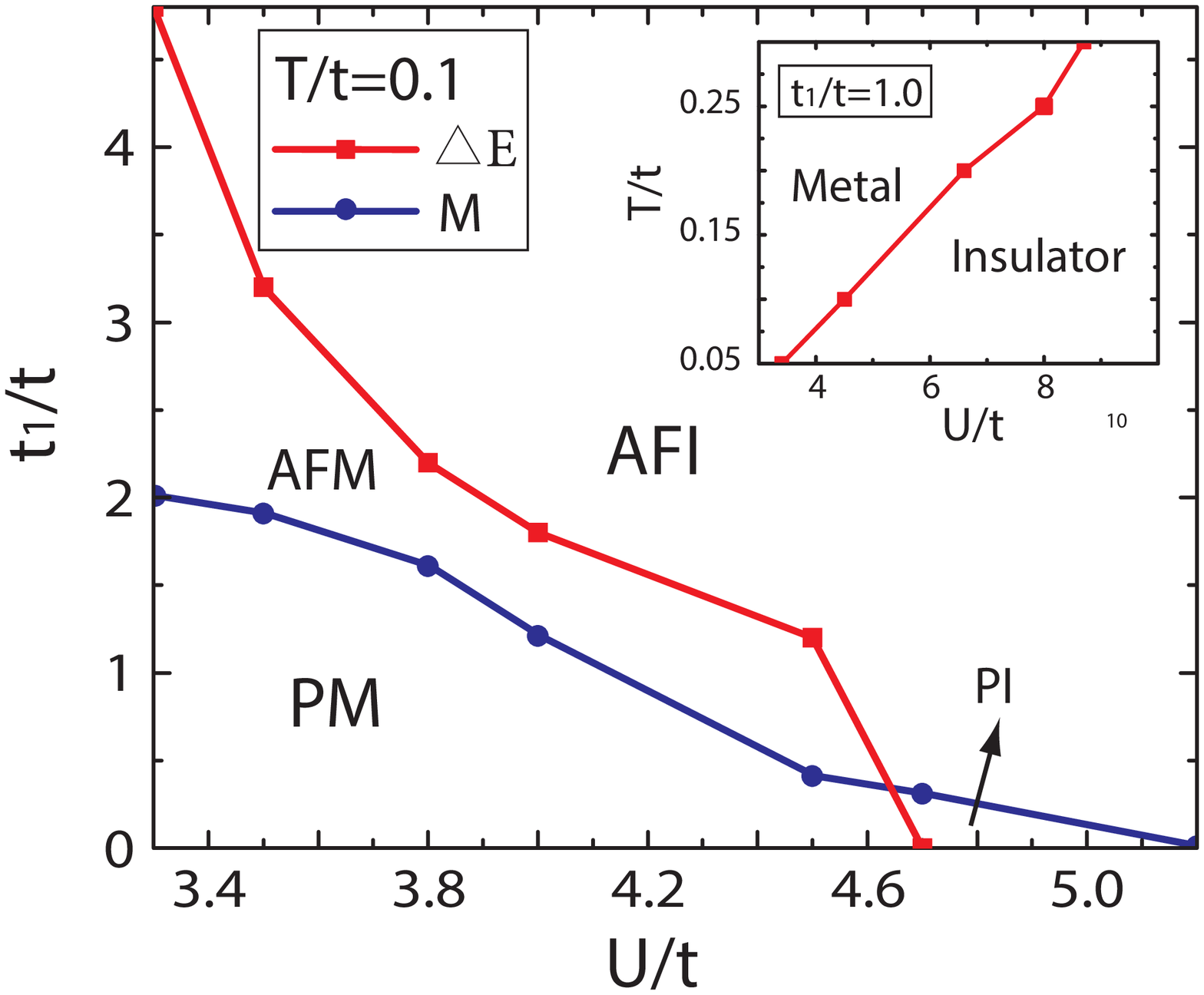,width=15cm}
\end{center}
\label{fig:PD}
\end{figure}
\begin{figure}
    \begin{center}
        \epsfig{file=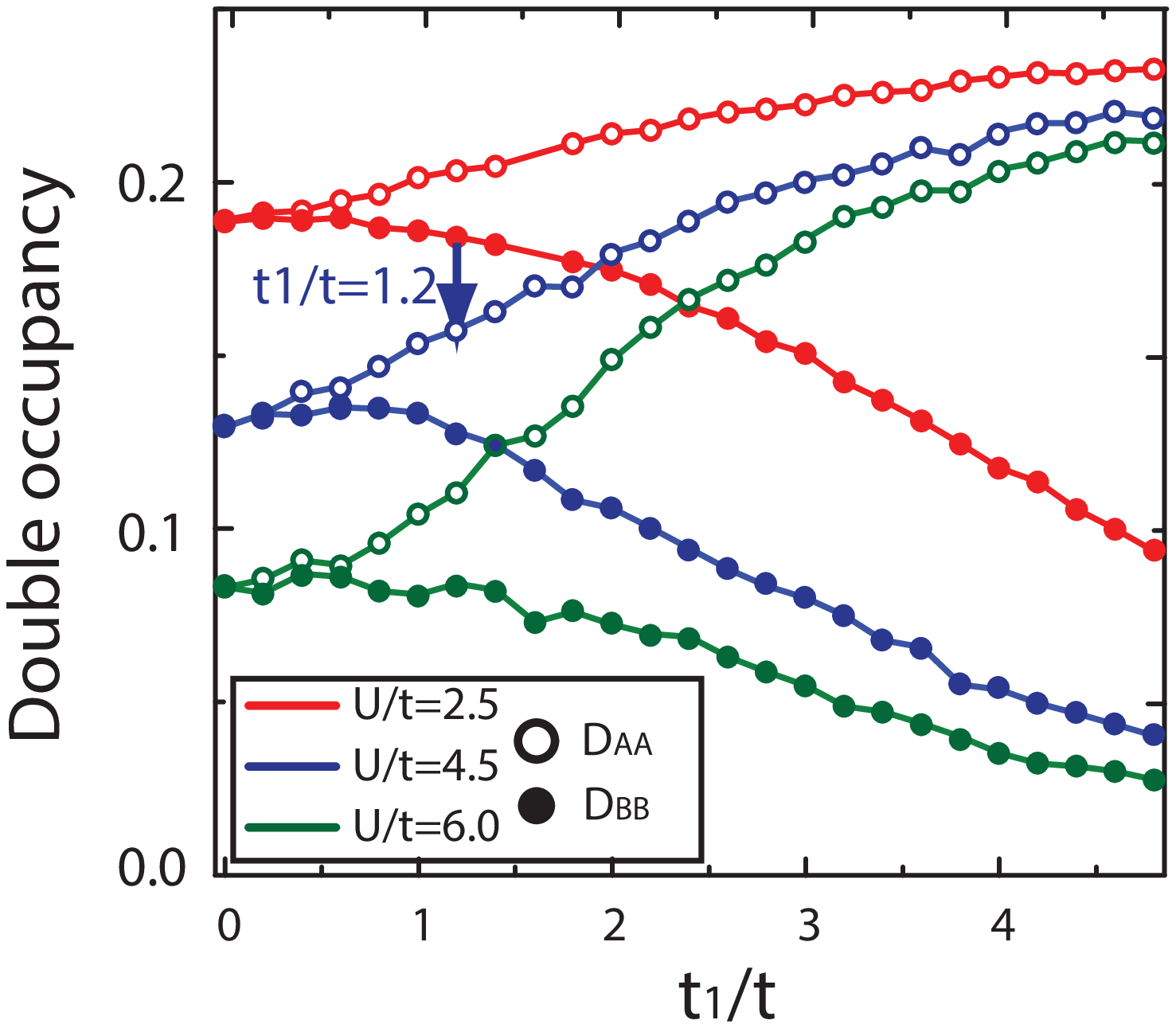,width=15cm}
    \end{center}
    \label{fig:Docc}
\end{figure}

\begin{figure}
    \begin{center}
        \epsfig{file=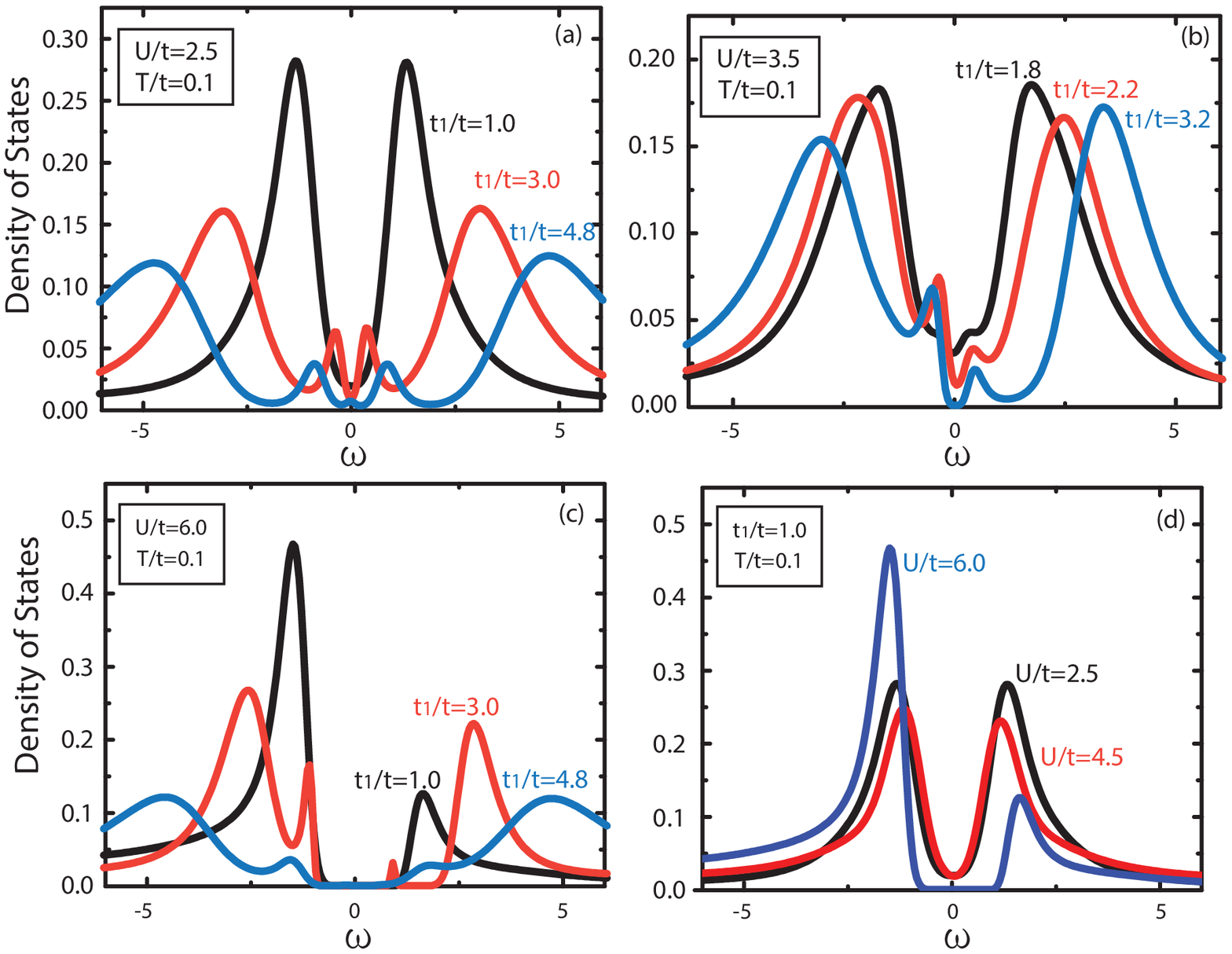,width=15cm}
    \end{center}
    \label{fig:DOS}
\end{figure}

\begin{figure}
    \begin{center}
        \epsfig{file=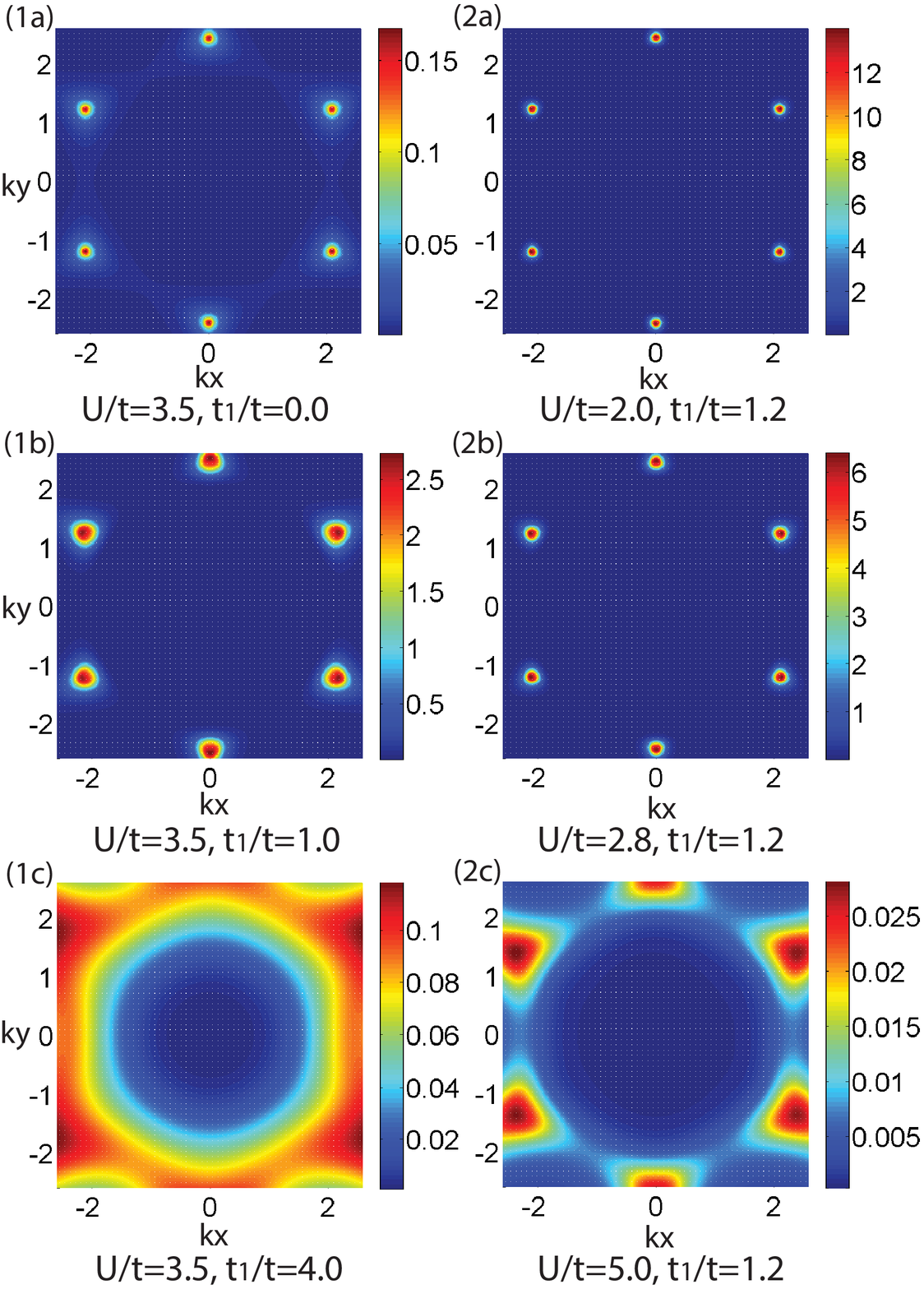,width=15cm}
    \end{center}
    \label{fig:FS}
\end{figure}

\begin{figure}
    \begin{center}
        \epsfig{file=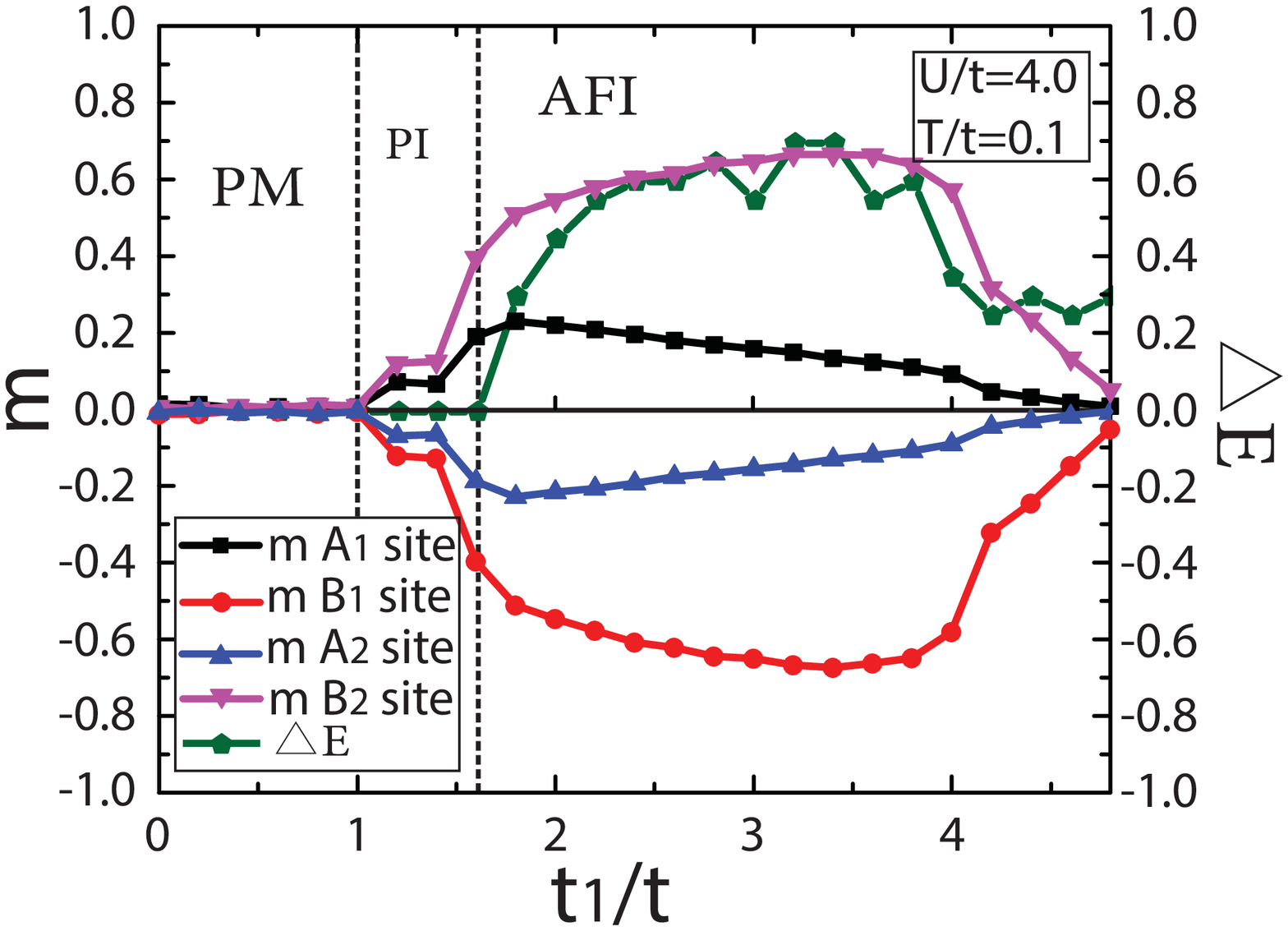,width=15cm}
    \end{center}
    \label{fig:magnetic}
\end{figure}

\begin{figure}
    \begin{center}
        \epsfig{file=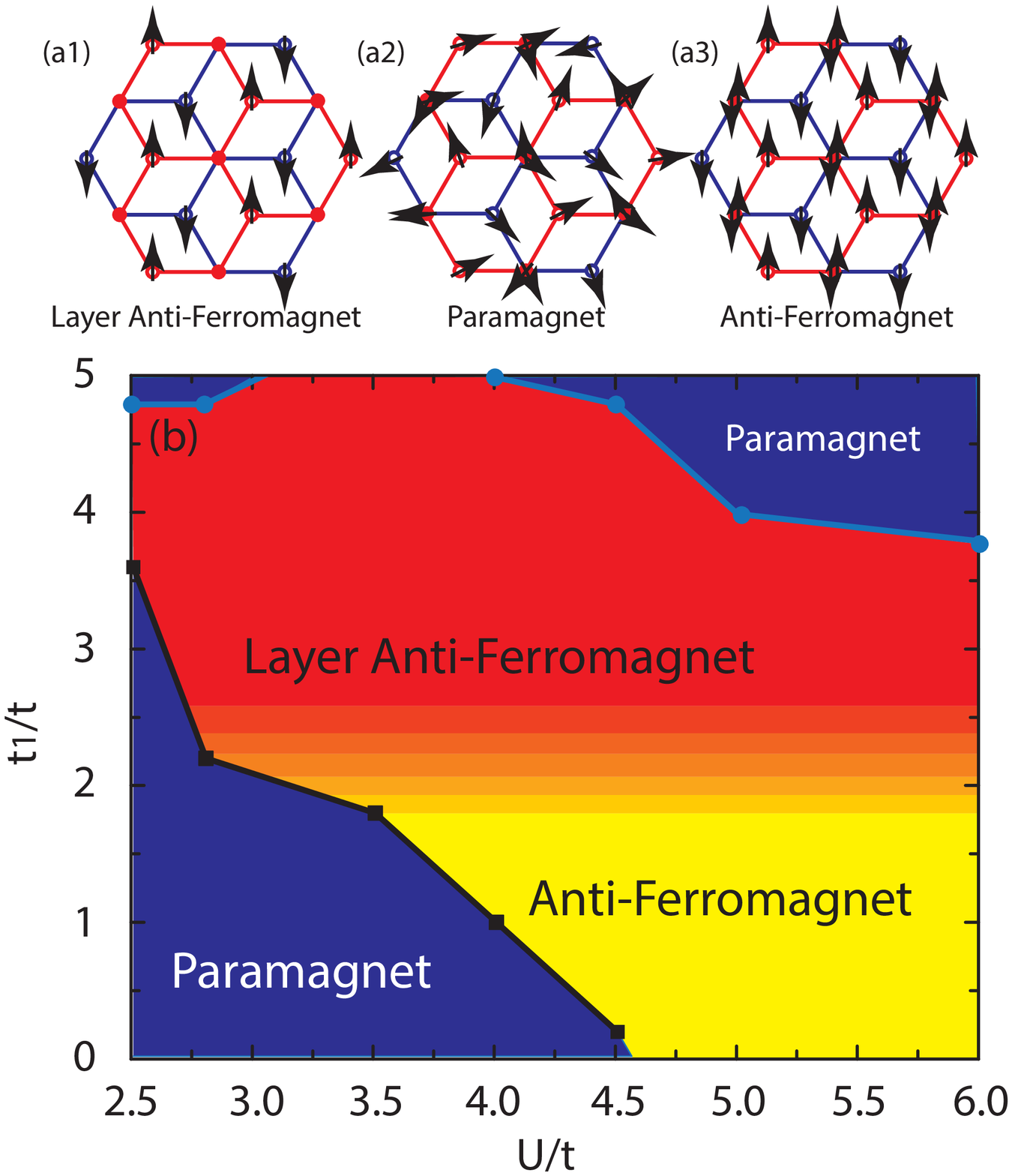,width=12cm}
    \end{center}
    \label{fig:PDmagnetic}
\end{figure}


\begin{thebibliography}{99}

\bibitem{Kharitonov} Kharitonov, M. Canted Antiferromagnetic Phase of $\nu=0$ Quantum Hall State in Bilayer Graphene. \textit{Phys. Rev. Lett.} \textbf{109,} 046803 (2012).
\bibitem{Novoselov} Novoselov, K. S., McCann, E., Morozov, S. V., Fal'ko, V. I., Katsnelson, M. I., Zeitler, U., Jiang, D., Shedin, F. \& Geim, A. K., Unconventional Quantum Hall Effect and Berry's Phase of $2\pi$ in Bilayer Graphene. \textit{Nature Phys.} \textbf{2,} 177 (2006).
\bibitem{Freitag} Freitag, F., Trbovic, J., Weiss, M. \& Sh$\ddot{o}$nenberger, C. Spontaneously Gapped Ground State in Suspended Bilayer Graphene. \textit{Phys. Rev. Lett.} \textbf{108,} 076602 (2012).
\bibitem{Feldman} Feldman, B. E., Martin, J. \& Yacoby, A. Broken-Symmetry States and Divergent Resistance in Suspended Bilayer Graphene. \textit{Nature Phys.} \textbf{5,} 889 (2009).
\bibitem{Zhao} Zhao, Y., Cadden-Zimansky, P., Jiang, Z. \& Kim, P. Symmetry Breaking in the Zero-Energy Landau Level in Bilayer Graphene. \textit{Phys. Rev. Lett.} \textbf{104,} 066801 (2010).
\bibitem{Weitz} Weitz, R. T., Allen, M. T., Feldman, B. E., Martin, J. \& Yacoby, A. Broken-Symmetry Sates in Doubly Gated Suspended Bilayer Graphene. \textit{Science} \textbf{330,} 812 (2010).
\bibitem{Maher} Maher, P., Dean, C. R., Young, A. F., Taniguchi, T., Shepard, K. L., Hone, J. \& Kim, P. Evidence for a Spin Phase Transition at Charge Neutrality in Bilayer Graphene. \textit{Nature Phys.} \textbf{9,} 154 (2013).
\bibitem{Kane} Kane, C. L. \& Mele, E. J. $Z_2$ Topological Order and Quantum Spin Hall Effect. \textit{Phys. Rev. Lett.} \textbf{95,} 146802 (2005).
\bibitem{Hosseini} Hosseini, M. V. \& Zareyan, M. Model of an Exotic Chiral Superconducting Phase in a Graphene Bilayer. \textit{Phys. Rev. Lett.} \textbf{108,} 147001 (2012).
\bibitem{Zhang} Zhang, Y. Y., Hu, J. P., Bernevig, B. A.,  Wang, X. R., Xie, X. C. \& Liu, W. M. Localization and the Kosterlitz-Thouless Transition in Disorderd Graphene. \textit{Phys. Rev. Lett.} \textbf{102,} 106401 (2009).
\bibitem{Mezzacapo} Mezzacapo, F. \& Boninsegni, M. Ground-State Phase Diagram of the Quantum $J_1-J_2$ model on the honeycomb lattice. \textit{Phys. Rev. B} \textbf{85,} 060402 (2012).
\bibitem{Anderson} Anderson, P. W. The Resonating Valence Bond State in $La_2CuO_4$ and Superconductivity. \textit{Science} \textbf{235,} 1196 (1987).
\bibitem{Wang} Wang M. et al. Antiferromagnetic Order and Superlattice Structure in Nonsuperconducting and Superconducting $Rb_yFe_{1.6+x}Se_2$ \textit{Phys. Rev. B} \textbf{84,} 094504 (2011).
\bibitem{Wu} Wu, W., Chen, Y. H., Tao, H. S., Tong, N. H. \& Liu, W. M. Interacting Dirac Fermions on Honeycomb Lattice. \textit{Phys. Rev. B} \textbf{82,} 245102 (2010)
\bibitem{Vafek} Vafek, O. Interacting Fermions on the Honeycomb Bilayer: From Weak to Strong Coupling. \textit{Phys. Rev. B} \textbf{82,} 205106 (2010).
\bibitem{McCann} McCann, E. \& Koshino, M. The Electronic Properties of Bilayer Graphene. \textit{Rep. Prog. Phys.} \textbf{76,} 056503 (2013).
\bibitem{Nilsson} Nilsson, J., Castro Neto, A. H., Peres, N. M. R. \& Guinea, F. Electron-Electron Interactions and the Phase Diagram of Graphene Bilayer. \textit{Phys. Rev. B} \textbf{73,} 214418 (2006).
\bibitem{McCann2} McCann, E. Asymmetry Gap in Electronic Band Structure of Bilayer Graphene. \textit{Phys. Rev. B} \textbf{74,} 161403 (2006).
\bibitem{Nilsson2} Nilsson, J., Castro Neto, A. H., Guinea, F. \& Peres, N. M. R. Electronic Properties of Graphene Multilayers. \textit{Phys. Rev. Lett.} \textbf{97,} 266801 (2006).
\bibitem{Lopes} Lopes dos Santos, J. M. B., Peres, N. M. R. \& Castro Neto, A. H. Graphene Bilayer with a Twist: Electronic Structure. \textit{Phys. Rev. Lett.} \textbf{99,} 256802 (2007).
\bibitem{Abergel} Abergel D. S. L. \& Chakraborty, T. Long-Range Coulomb Interaction in Bilayer Graphene. \textit{Phys. Rev. Lett.} \textbf{102,} 056807 (2009).
\bibitem{Jaksch} Jaksch, D., Bruder, C., Cirac, J. I., Gardiner, C. W. \& Zoller, P. Cold Bosonic Atoms in Optical Lattices. \textit{Phys. Rev. Lett.} \textbf{81,} 3108 (1998).
\bibitem{Hofstetter} Hofstetter, W., Cirac, J. I., Zoller, P., Demler, E. \& Lukin, M. D. High-Temperature Superfluidity of Fermionic Atoms in Optical Lattices. \textit{Phys. Rev. Lett.} \textbf{89,} 220407 (2002).
\bibitem{Greiner} Greiner, M., Mandel, O., Esslinger, T., H$\ddot{a}$nsch, T. W. \& Bloch, I. Quantum Phase Transition from a Superfluid to a Mott Insulator in a Gas of Ultracold Atoms. \textit{Nature {London}} \textbf{415,} 39 (2002).
\bibitem{Duan} Duan, L. M., Demler, E. \& Lukin, M. D. Controlling Spin Exchange Interactions of Ultracold Atoms in Optical Lattices. \textit{Phys. Rev. Lett.} \textbf{91,} 090402 (2003).
\bibitem{Soltan} Soltan-Panahi P. et al. Multi-Component Quantum Gases in Spin-Dependent Hexagonal Lattices. \textit{Nature Phys.} \textbf{7,} 434 (2011).
\bibitem{Gemelke} Gemelke, N., Zhang, X., Hung, C. -L. \& Chin, C. In Situ Observation of Incompressible Mott-Insulating Domains in Ultracold Atomic Gases. \textit{Nature (London)} \textbf{460,} 995 (2009).
\bibitem{Tao} Chen, Y. H., Tao, H. S., Yao, D. X. \& Liu, W. M. Kondo Metal and Ferrimagnetic Insulator on the Triangular Kagome Lattice. \textit{Phys. Rev. Lett.} \textbf{108,} 246402 (2012).
\bibitem{Metzner} Metzner, W. \& Vollhardt, D. Correlated Lattice Fermions $d=\infty$ Dimensions. \textit{Phys. Rev. Lett.} \textbf{62,} 324 (1989).
\bibitem{Georges1} Georges, A. \& Kotliar, G. Hubbard Model in Infinite Dimensions. \textit{Phys. Rev. B} \textbf{45,} 6479 (1992).
\bibitem{Bulla} Bulla, R. Zero Temperature Metal-Insulator Transition in the Infinite-Dimensional Hubbard Model. \textit{Phys. Rev. Lett.} \textbf{83,} 136 (1999).
\bibitem{Georges2} Georges, A. \&  Kotliar, G. Dynamical Mean-Field Theory of Strongly Correlated Fermion Systems and the Limit of Infinite Dimensions. \textit{Rev. Mod. Phys.} \textbf{68,} 13 (1996).
\bibitem{Kotliar} Kotliar, G., Savrasov, S. Y., P\'{a}lsson, G. \& Biroli, G. Cellular Dynamical Mean Field Approach to Strongly Correlated Systems. \textit{Phys. Rev. Lett.} \textbf{87,} 186401 (2001).
\bibitem{Maier} Maier, T., Jarrell, M., Pruschke, T. \& Hettler, M. H. Quantum Cluster Theories. \textit{Rev. Mod. Phys.} \textbf{77,} 1027 (2005).
\bibitem{Tong} Tong, N. H. Extended Variational Cluster Approximation for Correlated Systems. \textit{Phys. Rev. B} \textbf{72,} 115104 (2005).
\bibitem{Bolech} Bolech, C. J., Kancharla, S. S. \& Kotliar, G. Cellular Dynamical Mean-Field Theory for the One-Dimensional Extended Hubbard Model. \textit{Phys. Rev. B} \textbf{67,} 075110 (2003).
\bibitem{congjun} Cai, Z., Hung, H. H., Wang, L. \& Wu, C. J. Quantum Magnetic Properties of the $SU(2N)$ Hubbard Model in the Square Lattice: A Quantum Monte Carlo Study. \textit{Phys. Rev. B} \textbf{88,} 125108 (2013).
\bibitem{Rubtsov} Rubtsov, A. N., Savkin, V. V. \& Lichtenstein, A. I. Continuous-Time Quantum Monte Carlo Method for Fermions. \textit{Phys. Rev. B} \textbf{72,} 035122 (2005).
\bibitem{Zhu} Zhu, S. L., Wang, B., \& Duan, L. M. Simulation and Detection of Dirac Fermions with Cold Atoms in an Optical Lattice. \textit{Phys. Rev. Lett.} \textbf{98,} 260402 (2007).
\bibitem{Meng} Meng Z. Y. et al. Quantum Spin Liquid Emerging in Two-Dimensional Correlated Dirac Fermions. \textit{Nature (London)} \textbf{464,} 847 (2010).
\bibitem{Hohenadler} Hohenadler, M., Lang, T. C. \& Assaad, F. F. Correlation Effects in Quantum Spin-Hall Insulator: A Quantum Monte Carlo Study. \textit{Phys. Rev. Lett.} \textbf{106,} 100403 (2011).
\bibitem{Rachel} Wu, W., Rachel, S., Liu, W. M. \& Hur K. L. Quantum Spin Hall Insulator with Interactions and Lattice Anisotropy. \textit{Phys. Rev. B} \textbf{85,} 205102 (2012).
\bibitem{Kancharla} Kancharla S. S. \& Okamoto S. Band Insulator to Mott Insulator Transition in a Bilayer Hubbard Model. \textit{Phys. Rev. B} \textbf{75,} 193103 (2007).
\bibitem{Hu} Hu, H., Jiang, L., Liu, X. J. \& Pu H. Probing Anisotropic Superfluidity in Atomic Fermi Gases with Rashba Spin-Orbit Coupling. \textit{Phys. Rev. Lett.} \textbf{107,} 195304 (2011).
\bibitem{Jarrell} Jarrell, M. \&  Gubernatis, J. E. Bayesian Inference and the Analytic Continuation of Imaginary-Time Quantum Monte Carlo Data. \textit{Phys. Rep.} \textbf{269,} 133 (1996).
\bibitem{O'Hara} O'Hara, K. M., Hemmer, S. L., Gehm, M. E., Granade, S. R. \& Thomas, J. E. Observation of a Strongly Interacting Degenerate Fermi Gas of Atoms. \textit{Science} \textbf{298,} 2179 (2002).
\bibitem{Lucia} Hackerm\"{u}ller L. et al. Anomalous Expansion of Attractively Interacting Fermionic Atoms in an Optical Lattice. \textit{Science} \textbf{327,} 1621 (2010).
\bibitem{Tung} Tung, S., Schweikhard, V. \& Cornell, E. A. Observation of Vortex Pinning in Bose-Einstein Condensates. \textit{Phys. Rev. Lett.} \textbf{97,} 240402 (2006).
\bibitem{Hou} Hou, J. M. Energy Bands and Landau Levels of Ultracold Fermions in the Bilayer Honeycomb Optical Lattice. \textit{J. Mod. Opt.} \textbf{56,} 1182 (2009).
\bibitem{Schneider} Schneider, U., Hackermller, L., Will, S., Best, Th., Bloch, I., Costi, T. A., Helmes, R. W., Rasch, D. \& Rosch, A. Metallic and Insulating Phases of Repulsively Interacting Fermions in a 3D Optical Lattice. \textit{Science} \textbf{322,} 1520 (2009).
\bibitem{Jordans} J\"{o}rdans, R., Shohmaier, N., G\"{u}nter, K., Moritz, H. \& Esslinger, T. A Mott Insulator of Fermionic Atoms in an Optical Lattice. \textit{Nature (London)} \textbf{455,} 204(2008).
\bibitem{stoferle} S\"{o}ferle, T., Moritz, H., G\"{u}nter, K., K\"{o}hl, M. \& Esslinger, T. Molecules of Fermionic Atoms in an Optical Lattice. \textit{Phys. Rev. Lett.} \textbf{96,} 030401 (2006).
\bibitem{Kohl} K\"{o}hl, M., Moritz, H., St\"{o}ferle, T., G\"{u}nter, K. \& Esslinger, T. Fermionic Atoms in a Three Dimensional Optical Lattice: Observing Fermi Surface, Dynamics, and Interactions. \textit{Phys. Rev. Lett.} \textbf{94,} 080403 (2005).
\bibitem{Chin} Chin, J. K., Miller, D. E., Liu, Y., Stan, C., Setiawan, W., Sanner, C., Xu, K. \& Ketterle, W. Evidence for superfluidity of ultracold fermions in an optical lattice. \textit{Nature (London)} \textbf{443,} 961 (2006).
\bibitem{Hartmann} M\"{u}ller-Hartmann, E. The Hubbard Model at High Dimensions: Some Exact Results and Weak Coupling Theory. \textit{Z. Phys. B} \textbf{74,} 507 (1989).


\end{thebibliography}
\end{document}